\documentclass[trackchanges,twocolumn]{aastex7}
\usepackage{subcaption}
\usepackage{amsmath}

\defcitealias{2026ApJS..282....3S}{Sarrouh \& Asada et al.}
\newif\ifFIGSET
\FIGSETfalse
\begin{document}

\title{Interaction-induced star formation boosts stellar mass assembly in $z\sim5$ galaxies}

\author[orcid=0000-0002-8432-6870,sname='Omori']{Kiyoaki Christopher Omori}
\affiliation{Department of Astronomy and Physics and Institute for Computational Astrophysics, Saint Mary's University, 923 Robie Street, Halifax, Nova Scotia B3H 3C3, Canada}
\email{kiyoaki.omori@smu.ca}  

\author[orcid=0000-0002-7712-7857,sname='Sawicki']{Marcin Sawicki}
\affiliation{Department of Astronomy and Physics and Institute for Computational Astrophysics, Saint Mary's University, 923 Robie Street, Halifax, Nova Scotia B3H 3C3, Canada}
\email{marcin.sawicki@smu.ca}

\author[orcid=0000-0001-8115-5845,sname='M\'erida']{Rosa M. M\'erida}
\affiliation{Department of Astronomy and Physics and Institute for Computational Astrophysics, Saint Mary's University, 923 Robie Street, Halifax, Nova Scotia B3H 3C3, Canada}
\email{rosa.meridagonzalez@smu.ca}

\author[orcid=0000-0001-8325-1742,sname='Desprez']{Guillaume Desprez}
\affiliation{Department of Astronomy and Physics and Institute for Computational Astrophysics, Saint Mary's University, 923 Robie Street, Halifax, Nova Scotia B3H 3C3, Canada}
\affiliation{Kapteyn Astronomical Institute, University of Groningen, P.O. Box 800, 9700AV Groningen, The Netherlands}
\email{guillaume.desprez@protonmail.com}

\author[orcid=0000-0002-4542-921X, sname='Abraham']{Roberto Abraham} 
\affiliation{David A. Dunlap Department of Astronomy and Astrophysics, University of Toronto, 50 St. George Street, Toronto, Ontario, M5S 3H4, Canada}
\email{roberto.abraham@utoronto.ca}

\author[orcid=0000-0001-5984-0395, sname='Brada\v{c}']{Maru\v{s}a Brada\v{c}} 
\affiliation{Faculty of Mathematics and Physics, Jadranska ulica 19, SI-1000 Ljubljana, Slovenia}
\affiliation{Department of Physics and Astronomy, University of California Davis, 1 Shields Avenue, Davis, CA 95616, USA}
\email{marusa.bradac@fmf.uni-lj.si}

\author[orcid=0000-0003-3243-9969, sname='Martis']{Nicholas S. Martis} 
\affiliation{Faculty of Mathematics and Physics, Jadranska ulica 19, SI-1000 Ljubljana, Slovenia}
\email{nicholas.martis@fmf.uni-lj.si}

\author[orcid=0000-0002-9330-9108, sname='Muzzin']{Adam Muzzin} 
\affiliation{Department of Physics and Astronomy, York University, 4700 Keele St. Toronto, Ontario, M3J 1P3, Canada}
\email{muzzin@yorku.ca}

\author[sname='Noirot']{Ga\"el Noirot} 
\affiliation{Space Telescope Science Institute, 3700 San Martin Drive, Baltimore, Maryland 21218, USA}
\email{gnoirot@stsci.edu}

\author[orcid=0000-0001-8830-2166, sname='Sarrouh']{Ghassan T. E. Sarrouh} 
\affiliation{Department of Physics and Astronomy, York University, 4700 Keele St. Toronto, Ontario, M3J 1P3, Canada}
\email{gsarrouh@yorku.ca}

\author[orcid=0000-0002-4201-7367, sname='Willott']{Chris J. Willott} 
\affiliation{National Research Council of Canada, Herzberg Astronomy \& Astrophysics Research Centre, 5071 West Saanich Road, Victoria, BC, V9E 2E7, Canada}
\email{chris.willott@nrc.ca}

\author[orcid=0000-0001-9002-3502, sname='Marchesini']{Danilo Marchesini} 
\affiliation{Department of Physics and Astronomy, Tufts University, 574 Boston Avenue, Suite 304, Medford, MA 02155, USA}
\email{danilo.marchesini@tufts.edu}

\author[orcid=0009-0009-2307-2350, sname='Myers']{Katherine Myers} 
\affiliation{Department of Physics and Astronomy, York University, 4700 Keele St. Toronto, Ontario, M3J 1P3, Canada}
\email{kjmyers@yorku.ca}


\begin{abstract}

Galaxy interactions are a key ingredient in galaxy evolution; not only are they a primary pathway of galaxy growth and mass assembly, but also a key driver of processes such as star formation and quenching. We investigate the impact of galaxy-galaxy interactions on stellar mass assembly using JWST/NIRCam observations of a spectroscopically selected sample of galaxies at $5.0<z_{spec}<5.6$ from the Canadian NIRISS Unbiased Cluster Survey (CANUCS). Of the 48 galaxies in our parent sample, we visually classify 21 ($44\%$) as closely-interacting ($\lesssim$~5~kpc) system {candidates} with two or more components. We evaluate the non-parametric star formation histories (SFHs) of these systems' components using the spectral energy distribution fitting code \textsc{Dense Basis}. We find that the components in these systems experience brief intervals ($\sim0.2$ Gyr) of strongly enhanced star formation that grow their stellar mass by $\sim2.66\pm0.85\times$, forming $\sim1.58\pm0.25\times$ of excess mass than expected compared to if there was no burst. Attributing these star formation rate enhancements to interactions and assuming that the components will merge, we find that mergers are responsible for $\sim38^{+18}_{-20}\%$ of the total stellar mass growth of galaxies at $z\sim5$. While {there is significant} contribution {to mass growth} from the merging of the pre-existing stellar masses of the merging galaxies, {up to }half {is} due to stellar mass that is newly-formed during the interaction. We conclude that mergers, and their associated star formation bursts, are an important pathway for stellar mass growth in high-$z$ galaxies.

\end{abstract}



\section{Introduction} \label{sec:intro}

In the currently accepted framework for structure formation in the Universe, hierarchical merging is considered to be a key process in galaxy evolution \citep{2000MNRAS.319..168C}. During galaxy interactions and mergers, strong gravitational torques, the dissipation of angular momentum, and the subsequent infall of gas towards the central regions of galaxies can trigger strong star formation activity and starbursts \citep{2000ApJ...530..660B,2005ASSL..329..143S,2009ApJ...694L.123K,2009PASJ...61..481S,2014MNRAS.442L..33R,2015MNRAS.446.2038R,2019MNRAS.490.2139R}. This interaction-induced star formation can produce additional stars that increase the stellar mass of the resulting post-merger galaxy beyond what it would have accrued simply from combining the stellar masses of the progenitors.


A large number of studies have shown that in the local universe star formation can be enhanced in merging systems compared to isolated galaxies \citep{2008ChJAS...8...77B, 2008AJ....135.1877E,2011MNRAS.412..591P,2013MNRAS.433L..59P,2013MNRAS.430.1901H,2015MNRAS.448.1107M,2016MNRAS.462.2418S,2019MNRAS.482L..55T, 2021ApJ...923..205Y,2024MNRAS.527.6506B}. While the reported degree of enhancement differs based on factors such as sample selection and property measurement, in the local universe merger-induced star formation contributes to excess stellar mass growth, with enhancement factors ranging from $10\%$ to $ 20\%$ \citep{2025MNRAS.538L..31F} over the mass growth from the combination of the pre-existing stellar populations. {Notably, interactions produce a boost in the star formation rate that is relatively short-lived and consequently {most} detectable when the interacting galaxies are at the smallest separations ({$\lesssim 30$~kpc}). The enhancement becomes less pronounced as pair separation increases \citep{2008AJ....135.1877E, 2010MNRAS.407.1514E,2011MNRAS.412..591P,2013MNRAS.433L..59P,2016MNRAS.461.2589P,2012MNRAS.426..549S,2015MNRAS.452..616D,2020MNRAS.494.4969P}. However, other works do not detect enhancements even at separations of $\sim10$ kpc \citep{2018ApJ...868...46S,2026A&A...706A.264C}, and the orbital extent of influence of enhanced star formation is not easy to discern.}
At high redshifts ($z\sim5$ and above), the merger fraction has been found to be much higher than at lower redshifts, increasing monotonically with redshift and ranging between $\sim20-40\%$ \citep{2017A&A...608A...9V,2019ApJ...876..110D,2024MNRAS.52711372A,2025MNRAS.544.1412S,2025MNRAS.540..774D, 2025MNRAS.540.2146P}. 
Analyzing samples of high-$z$ galaxy pairs \citet{2025MNRAS.540..774D, 2025MNRAS.540.2146P} {found that while mergers can enhance the star formation activity in a galaxy, their contribution to stellar mass growth in galaxies are smaller compared to intrinsic star formation.}
However, these works focused largely on widely-separated pairs, potentially missing the star formation enhancement which is {most pronounced at small separations ($<10$ kpc) even at low redshifts \citep{2025MNRAS.538L..31F,2025ApJ...989...73O}.}
 
For high-$z$ pairs at small separations, \citet{2025arXiv251014743P} find a moderate enhancement for longer timescale ($\sim50-100$ Myr) star formation rates, but find that shorter timescale ($\sim5-10$ Myr) star formation is dominated by internal processes.
The analysis of star formation burstiness presented by \citet{2024MNRAS.52711372A} indicates that interactions may indeed play a role in regulating star formation. These authors found that while both close galaxy pairs (projected separations of $\lesssim5$kpc) and isolated galaxies experience bursty star formation, the bursts in closely paired galaxies quench much more rapidly than those in isolated galaxies. This observed dichotomy suggests that star formation in close pairs may be associated with a different mode of star formation than the presumably stochastic busts in isolated systems, and galaxy-galaxy interactions are an obvious culprit.

In this paper we examine the star formation histories and the enhancement of star formation in $z\sim5$ interacting systems with small intercomponent separations. 
Since \citet{2024MNRAS.52711372A} have already demonstrated that the star formation histories (SFHs) of isolated galaxies differ systematically from those of close galaxy pairs at these redshifts, our goal here is not to revisit the comparison between paired and isolated systems, but rather to investigate the nature and stellar-mass consequences of the star formation bursts occurring within the close-pair population itself.
To do so, we use Near Infrared Camera (NIRCam; \citealt{2023PASP..135b8001R}) images of galaxies with Near Infrared Spectrograph (NIRSpec; \citealt{2022A&A...661A..80J}) spectroscopic redshifts in the range $5<z_{spec}<5.6$ that were observed as part of the JWST GTO program Canadian NIRISS Unbiased Cluster Survey (CANUCS, GTO: 1208, \citealt{2022PASP..134b5002W}). Galaxies in this spectroscopic redshift range are guaranteed to have their H$\alpha$ emission line within the NIRCam F410M medium band filter and this allows us to map the recent short timescale ($\sim$10~Myr) star formation across the interacting system irrespective of NIRSpec microshutter placement geometry. The reliance on spectroscopic redshifts for the selection of the sample biases our sample in that it requires our galaxies to have been spectroscopically targeted and then successfully detected; however, it allows us to confidently evaluate the short-timescale star formation histories of all the components within each system. 



Throughout, we adopt a $\Lambda$CDM cosmological model with $H_0 = 70\textrm{ km } \textrm{s }^{-1} \textrm{ Mpc}^{-1}$, $\Omega_{\Lambda}=0.7$, $\Omega_{M}=0.3$. 
In this cosmology, 1 arsecond corresponds to $5.4-6.0$ physical kpc within our target redshift range of $z=5.0-5.6$.
We use AB magnitudes \citep{1983ApJ...266..713O} and 
assume the Chabrier \citep{2003PASP..115..763C} stellar initial mass function (IMF) when deriving physical properties through photometric SED fitting.

\section{Data} \label{sec:highlight}

\subsection{CANUCS imaging and spectroscopy}

This work uses observational data obtained as part of the CANUCS survey, which offers a combination of NIRCam imaging, NIRISS slitless spectroscopy, and NIRSpec prism multi-object spectroscopy reaching a limiting magnitude of $\sim$30 AB mag at a 5$\sigma$ detection threshold (\citetalias{2026ApJS..282....3S}  2026). CANUCS observed five gravitationally lensing clusters, Abell370, MACS0416, MACS1149, MACS0417, and MACS1423. From each of these, we use imaging data obtained in two  pointings: the cluster center field (hereafter CLU), and flanking fields (hereafter NCF). Depending on cluster and field, observations were made using a combination of the NIRCam and NIRISS broadband filters F090W, F115W, F150W, F200W, F277W, F356W, and F444W and medium band filters F140M, F162M, F182M, F210M, F250M, F300M, F360M, F410M. Additional observations were made on three of the clusters (Abell370, MACS0416, MACS1149) in the Cycle 2 ``JWST in Technicolor” program (ID: 3362, PI: A. Muzzin), where 8 wide, medium and narrow band filters were added (F070W, F164N, F187N, F200W, F356W, F430M, F460M, and F480M) for the NCF fields and F090WN for the CLU fields. 

Supplementary \textit{HST} data are also available in these fields. In the CLU fields of Abell370, MACS0416, MACS1149, ultra-deep imaging in the optical wavelengths are available with \textit{Hubble Space Telescope} (\textit{HST}) Advanced Camera for Surveys (ACS) in F435W, F606W, and F814W, and in the infrared with \textit{HST} Wide Field Camera 3 (WFC3/IR) in F105W, F125W, F140W, F160W \citep{2017ApJ...837...97L}, as well as WFC3/IR F110W \citep{2012ApJS..199...25P}. The NCF fields of  Abell370, MACS0416, and MACS1149 are supplemented by similar \textit{HST}/ACS data, and MACS0417 and MACS1423 are supplemented with \textit{HST}/WFC3 Ultraviolet Imaging Spectrograph (UVIS) imaging in the F438W and F606W bands (HST-GO16667; PI: M.Bradac). {Each filter also has root mean squared (RMS) uncertainty maps available, where the noise level in each pixel is estimated by rescaling the corresponding weight image such that the sky background of the RMS-normalized science image follows a standard normal distribution.}

CANUCS also consists of NIRSpec spectroscopy, primarily in the CLU fields but with some in the NCF pointings. This spectroscopy was conducted as follow-up on a diverse set of targets of interest selected from the NIRCam, NIRISS, and \textit{HST} Imaging using a mixture of selection criteria (see \citetalias{2026ApJS..282....3S} 2026).  The targets were observed using the low-resolution ($R\sim100$) prism, covering the wavelength range from $0.6-5.5\mu$m. Observations were done with three MSA configurations in each of the five CLU fields, with the exception of MACS0417, which used two configurations in the CLU field and one in the NCF field. Each configuration was observed for 2.9 ks, with a small set of high-priority targets receiving up to 8.7 ks exposure time.

We use physical properties, such as stellar mass ($M_*$) and star formation rate (SFR), measured using the SED fitting code \textsc{Dense Basis} \citep{2017ApJ...838..127I,2019ApJ...879..116I}, which are available for all galaxies in the CANUCS Data Release database ( \citetalias{2026ApJS..282....3S} 2026). Further details regarding CANUCS/Technicolor data reduction, calibration, and data products are available in \citet{2022PASP..134b5002W} and 
\citetalias{2026ApJS..282....3S} (2026).

\subsection{Our sample}

Among the galaxies  in the CANUCS spectroscopic catalog, we have a mixed sample of galaxies located in our target redshift range of $5.0 < z_{spec} \leq 5.6$. After eliminating two sources with morphological AGN signatures, our parent sample has 48 galaxies, of which we subsequently (Sec.~\ref{sec:GalaxyClassification}) identify 21 as having two or more closely-separated ($\lesssim$5 kpc) components reminiscent of interacting pairs or multiples. 

We study spectroscopically-confirmed galaxies in the $5.0 < z_{spec} \leq 5.6$ redshift range as the H$\alpha$ emission, which we can use as a tracer for short timescale star formation, falls in the NIRCam F410M filter. While a larger, more complete, and more representative sample of interacting systems could be selected using photometric redshifts (e.g., the approach used in \citealt{2024MNRAS.52711372A}), in this study we opted to use only galaxies with spectroscopic redshifts to avoid potential redshift-line flux degeneracies that could bias the results and to ensure that the SED-fitting results are robust.  

We stress that the requirement for spectroscopy, particularly as CANUCS spectroscopic targets are selected through a mixture of selection criteria, introduces a set of potential biases. We thus eschew in this study comparisons between interacting galaxies and isolated ones, as such comparisons would likely be biased. We focus instead only on an assessment of star formation rate enhancements in the systems we deem to be interacting (see Sec.~\ref{sec:GalaxyClassification}). We acknowledge that this sample is likely biased towards line emitters that aid spectroscopic redshift determination; we deem this to be an acceptable trade-off to ensure the robustness of the SED-fitting and leave the more detailed analysis of a photometrically-selected sample to forthcoming work (K.\ C.\ Omori et al.,~in prep.).

\begin{figure*}[htbp]
    \centering
    \begin{subfigure}[b]{0.32\textwidth}
        \centering
        \includegraphics[width=\linewidth]{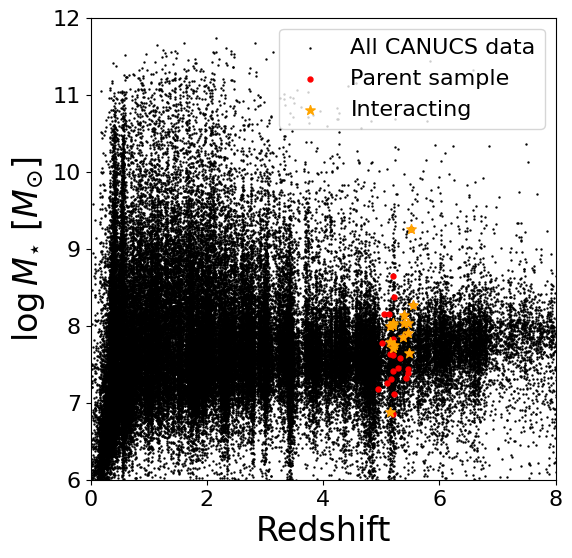}
        \label{fig:first}
    \end{subfigure}\hfill
    \begin{subfigure}[b]{0.32\textwidth}
        \centering
        \includegraphics[width=\linewidth]{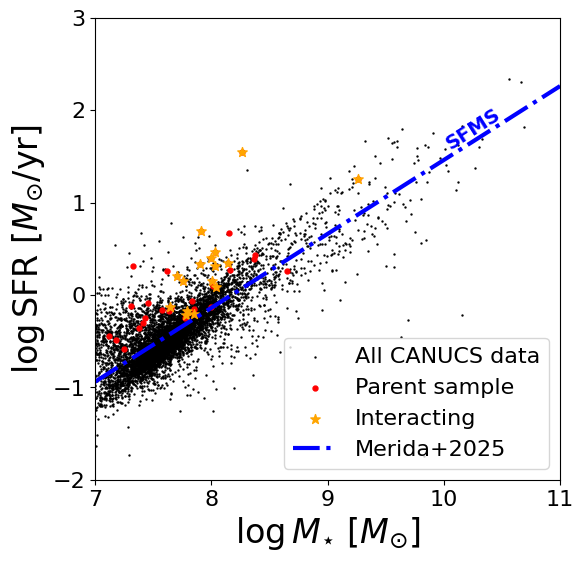}
        \label{fig:second}
    \end{subfigure}\hfill
    \begin{subfigure}[b]{0.3135\textwidth}
        \centering
        \includegraphics[width=\linewidth]{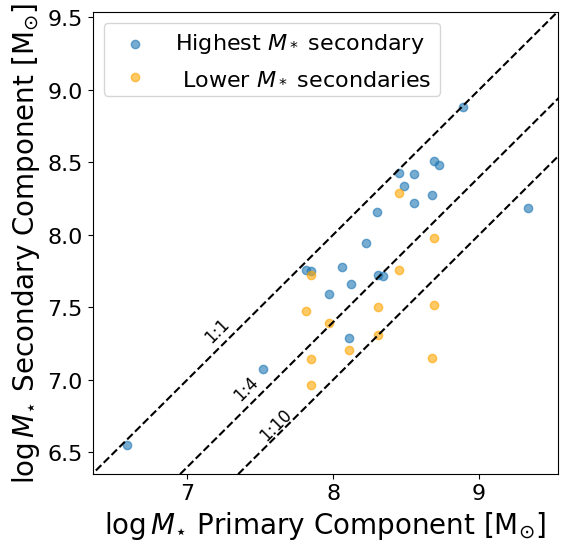}
        \label{fig:third}
    \end{subfigure}
    \hfill
    \caption{{Left:} redshift-$M_{*}$ distribution of our sample (red dots) and our interacting systems (orange stars), compared to the entire CANUCS catalog (black dots). {Middle:} $M_{*}$-SFR (over the last 100 Myr) distribution of our sample (red dots) and our interacting systems (orange stars), compared to galaxies in the CANUCS catalog between $5.0 < z \leq 5.6$ (black dots). The dash dotted blue line shows the star-forming main sequence (SFMS) at this redshift found in \citet{2025arXiv250922871M}. The majority of our interacting systems lie above the SFMS. {Right:} Distributions of the secondary component stellar masses as a function of the primary (most massive) component stellar mass.  Blue points compare the mass of the second highest mass (highest mass secondary) component relative to the highest mass (primary) component of each system. Yellow points compare the mass of the lower mass secondary components relative to the primary component. The dashed black lines indicate the boundary between 1:1 equal mass ratio and the 1:4 mass ratio threshold for major mergers, and the boundary between 1:4 mass ratio and 1:10 mass ratio for minor mergers. The majority of the highest-mass secondary components lie between the 1:1 and 1:4 mass ratio lines, suggesting that they are major mergers.}
    \label{fig:ms}
\end{figure*}

With this caveat in mind, Fig.~\ref{fig:ms} shows the redshift-$M_{*}$ and $M_{*}$-SFR properties of our spectroscopic parent sample in reference to galaxies in the CANUCS catalog with all properties, derived by \textsc{dense basis}, taken from the CANUCS database (\citetalias{2026ApJS..282....3S} 2026).
We find that both our parent sample and the sample of 21 multi-component systems are a broad representation of CANUCS galaxies in this redshift range in terms of $M_*$; the majority of them have elevated SFRs in reference to the main sequence found in \citet{2025arXiv250922871M}, as expected given the requirement for spectroscopic detection via the presence of emission lines.  We thus appear to be selecting objects that are caught during episodes of elevated star formation. 

\section{Methods} \label{sec:Data}


\subsection{Galaxy classification}\label{sec:GalaxyClassification}

For this study, we conduct visual classification based on galaxy RGB $40\times40$ postage stamp images made from combining F410M (R),
F277W (G), F150W and F200W (combined for B) {common PSF-matched} NIRCam imaging, and their segmentation maps (\citetalias{2026ApJS..282....3S} 2026). While visual classification is inefficient for large samples, it still remains a widely used method for merger identification of high redshift galaxies \citep{2024MNRAS.533.4472D,2024MNRAS.52711372A}.
Human-based classifications can also distinguish between intricate merger properties such as merger stage or progenitor morphology, and make purified subsamples on mergers classified using other methods 
\citep{2012A&A...539A..45L,2012MNRAS.419.2703H,2025ApJ...989...73O,2025MNRAS.tmp.1542D}.
As our identification is focused on a small parent sample from which we aim to select systems with close components, visual classification becomes the ideal method.

We classify systems in our parent sample to be interacting if they have two or more distinct close components that can be resolved in NIRCam RGB imaging. The multicolor image lets us incorporate color information into the classification. We confirm the results of this visual classification by inspecting the segmentation maps produced by CANUCS on the $\chi_{mean}$ detection image, {drizzled to a common pixel scale of 0.04 arcsecs per pixel}, that combines all available bands (see \citetalias{2026ApJS..282....3S} 2026 for details).
Using the CANUCS segmentation map as a mask, we deblend the detection image into components using \textsc{photutils} \citep{larry_bradley_2025_14889440}, controlled by the parameters \textsc{npixels} (minimum number of connected pixels) set to 3, \textsc{nlevels} (number of multi-thresholding levels) set to 64, and \textsc{contrast} (fraction of local peak flux with respect to total flux of object) set to 0.0001. 
After the \textsc{photutils} deblending, we visually inspect the segmentation maps, and remove any components with fewer than 5 pixels, {corresponding to 0.2 arcsecs}, as these are likely noise not associated with the galaxy.
If a galaxy that we visually classified as having multiple components is also deblended into multiple components in our \textsc{photutils} segmentation map, we consider it to be a `multi-component system'. 

\begin{figure*}[htbp]
    \centering
    \includegraphics[width=\linewidth]{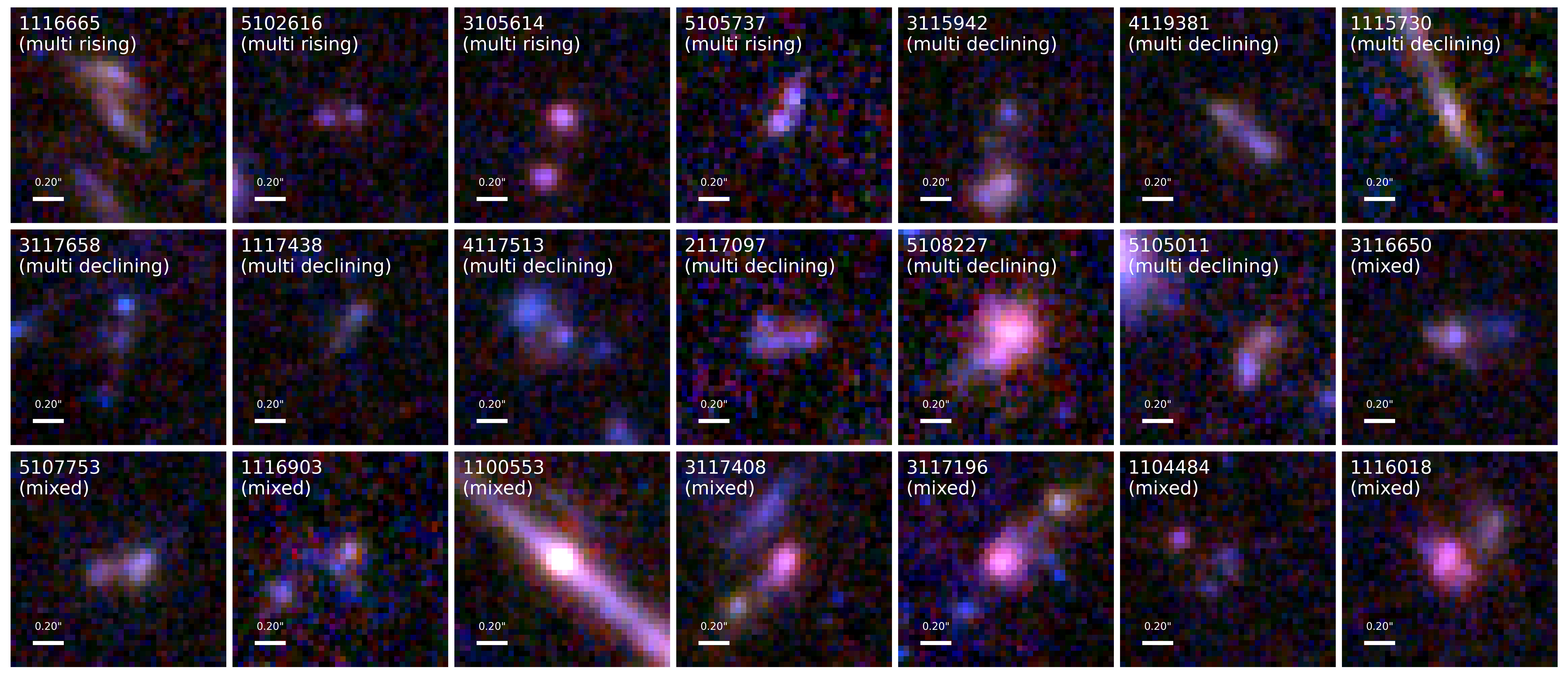}
    
    \vspace{0.5cm} 
    
    \includegraphics[width=\linewidth]{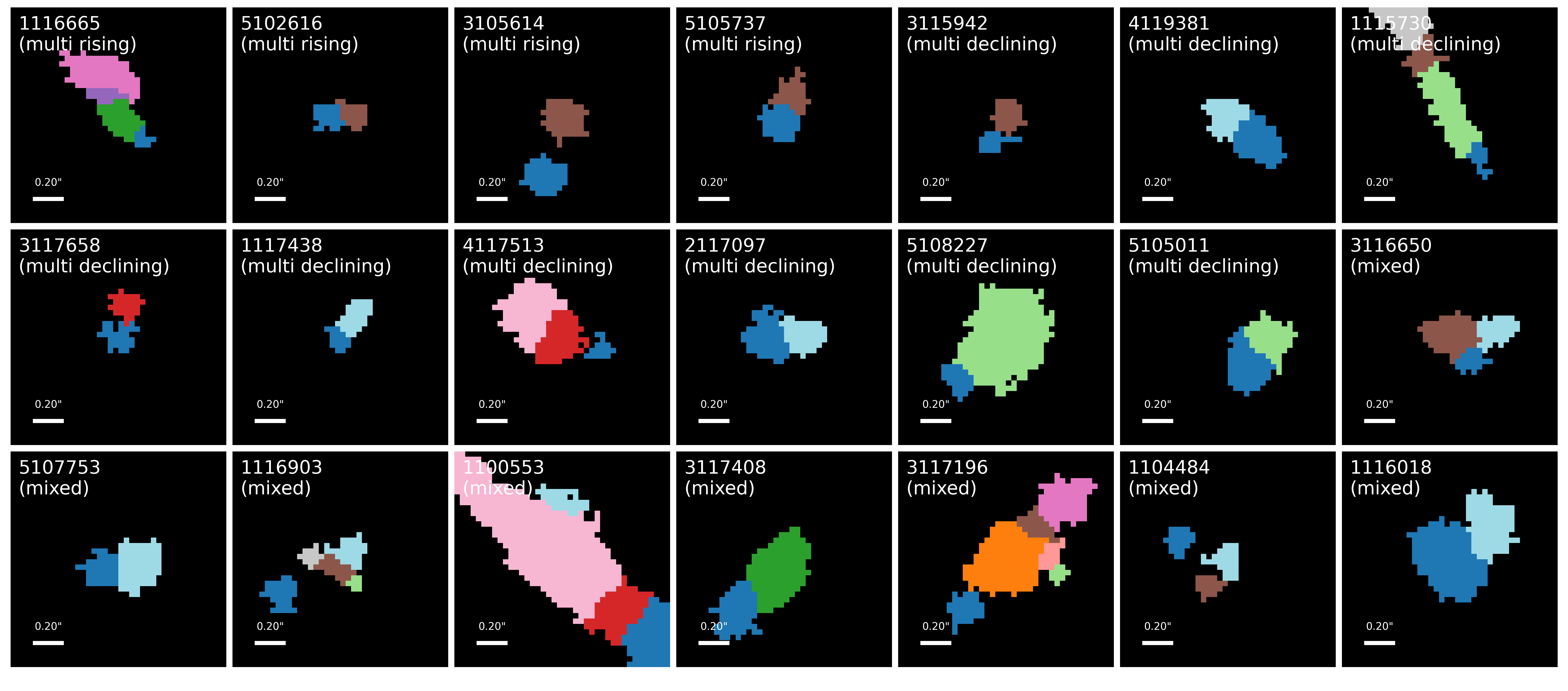}
    
    \caption{RGB image mosaic (top) and corresponding segmentation map mosaic (bottom) for the 21 interacting systems in our sample. Each cutout is centered on the spectroscopic target. RGB images are made by combining F410M (R), F277W (G), F150W and F200W (combined for B) NIRCam imaging{, and are drizzled to a common resolution of 0.04 arcsec per pixel}. The CANUCS source ID, and the SFH behaviours are indicated in the top left corner of each cutout. The image scale bar is plotted in the bottom left corner. The 0.2" scale bar corresponds to $\sim$1.2 physical kpc at the redshifts of these objects (and smaller still in cases of significant magnification), so these systems have components with very small separations.}
    \label{fig:combined_mosaics}
\end{figure*}

While the spectroscopic redshifts of the systems are not available for all components, we assume that the components within these multi-component systems are at a common redshift and interacting, which is plausible given their small separations on the sky. We nevertheless confirm that these multi-component systems are  not chance projections through evaluation of their photometric redshifts in conjunction with morphological disturbances. Photometric redshifts for each component were derived by running \textsc{EAzY} \citet{2021zndo...5012704B}, using the standard templates augmented with the \citet{2023ApJ...958..141L} set and the intergalactic medium attenuation curve of \citet{2025ApJ...983L...2A}. Through this step, we removed likely low redshift interlopers, such as the bottom component in ID 3115942 (row 1, column 5 in Fig. \ref{fig:combined_mosaics}), and the upper components in ID 3117408 (row 3, column 4 in Fig. \ref{fig:combined_mosaics}). We note that even if we remove all components where there is a major disagreement between the spectroscopic redshift and \textsc{EAzY} photometric redshift ($|\frac{z_{\mathrm{spec}-z_\mathrm{phot}}}{1+z_\mathrm{spec}}|>0.1$, {which accounts for roughly a $10\%$ discrepancy}), our quantitative results in the following sections do not change significantly.

Of the 48 spectroscopically selected systems in the parent sample, 21 were identified as interacting through our multi-step visual identification process, having a total of 58 distinct components. The other 27 systems were considered `single-component', where either one or both of the RGB image inspection or segmentation map showed the galaxies to be unresolved into distinct components, and there were no close companions in the postage stamp within 0.5 arcsec {($\sim3.2$ kpc)}. In this paper, we focus solely on the multi-component systems (for a comparison of star formation histories in interacting and non-interacting galaxies, see \citealt{2024MNRAS.52711372A}). 

We show our 21 multi-component systems and their segmentation maps in Fig. \ref{fig:combined_mosaics}.  As that figure shows, most of these systems are resolved into multiple, distinct, closely-spaced components both in their RGB images and in the segmentation maps. A few appear to be arcs with significant lensing magnification. Indeed, two highly-magnified systems that were studied individually in previous works are among our sample, namely  ID 1116665 \citep[row 1, column 1; called ELG1 and ELG2 by][]{2023MNRAS.523L..40A} and ID 1100553 \citep[row 3, column 3; dubbed z5BBG by][]{2023ApJ...949L..23S}. However, the majority of the systems in Fig.~\ref{fig:combined_mosaics} do not show signatures of strong distortion, with a median lensing magnification factor of $\mu\sim2.14$, and all have multiple components with separations of just a few tenths of arcsec. Given that 0.2~arcsec (the size of the scale bars shown in the Figure) at the redshifts spanned by our sample correspond to $\sim$1.2 kpc, these components are tightly grouped together within each system. {While the actual separations of components can be affected by differential magnification, the separations between our components should be an upper limit.}

We note that there are some caveats with our selection process and discuss them in Sec. \ref{caveats}.


\subsection{Component properties from SED fitting}
\label{SEDFIT}

Having segmented the multi-component systems into their individual components (Sec.~\ref{sec:GalaxyClassification}), we obtain fluxes of each component by summing their pixel fluxes. {Flux uncertainties are obtained through error propagation using the RMS maps. While CANUCS sources conduct corrections on fluxes and flux uncertainties accounting for apertures and inter-pixel correlations due to drizzling (see \citetalias{2026ApJS..282....3S} 2026 for details), the effects of these corrections are minimal towards our final results, as such we do not conduct these corrections in this work.} We then run the SED fitting code \textsc{Dense Basis} to obtain their physical properties, focusing in this paper on non-parametric SFHs, i.e., SFR(t).

SED-fitting methods that use non-parametric SFH methods, such as \textsc{Dense Basis}, do not assume a simple SFH form; rather, {they can flexibly describe a wide range of complex SFH shapes \citep{2017ApJ...838..127I,2019ApJ...879..116I,2019ApJ...876....3L,2021ApJS..254...22J,2022ApJ...927..170T,2026arXiv260521599D}}. {Non-parametric methods can also mitigate the problems of `outshining': an effect where emission from newly born stars outshines or masks those from older stars, making determination of stellar masses challenging \citep{1998AJ....115.1329S, 2018MNRAS.476.1532S, 2019ApJ...876....3L,2020ApJ...904...33L,2024ApJ...961...73N,2025MNRAS.542.2998H,2025ApJ...987..184W}}.

{\textsc{Dense Basis}-recovered SFHs are described using a tuple \{$M_*$, $SFR$, {$t_{\chi}$}\}, where $M_*$ is stellar mass, $SFR$ is SFR at the epoch of observation, and {$t_{\chi}$} contains a set of \textit{N} ``shape" parameters that describe mass-assembly times in the SFH. These are are \textit{N} lookback times at which the component formed equally spaced quantiles of its total mass. For this work, we set $N=3$, giving {$t_{25},t_{50},t_{75}$}, corresponding to the lookback times where the component formed $25\%,50\%,75\%$ of its stellar mass, respectively. This gives $N+1$ distinct mass-assembly time intervals. The shape of the SFH is determined by a Dirichlet prior $\alpha$, with further details on this prior available in \citet{2017ApJ...837..170L,2019ApJ...876....3L}. $\alpha$ can be set to a single value across the entire SFH, but also can be set individually to each mass-assembly time interval for fine-grained control. For this work, we set $\alpha=3.0$ as a baseline, and set $\alpha=18.0$ for the oldest mass-assembly time interval. This configuration allows for a smoother SFH at the earliest epochs while retaining a higher flexibility to capture complex SFH forms with bursts and recent quenching.}


We run \textsc{Dense Basis} on the photometric {fluxes and uncertainties} of the individual components assuming the Calzetti law \citep{2000ApJ...533..682C} for dust attenuation, with the following priors{, following the high redshift priors used in \citet{2024MNRAS.52711372A}}: redshifts fixed to the confirmed spectroscopic redshifts of each system; a uniform stellar mass prior of $5.0 < \log (M_{*}) < 10.0$; a flat specific star formation rate (sSFR) prior between {$-6.5 > \log(\mathrm{sSFR}) 
>-12$;} a flat metallicity prior between $-2.5 < \log(Z/Z_{\odot}) < -1.0$; and an exponential dust extinction ($A_V$) prior with a scale value of $A_V$ = 3.0 mag. {The metallicity prior is restricted to the values indicated to follow the metallicity of low-mass, high redshift galaxies studied in \citet{2023ApJS..269...33N,2024A&A...684A..75C}.} {As for the dust prior,} we also tested a flat dust prior but found that using it did not materially change the results. Because in the rest of the paper we focus on normalized star formation histories, which are not affected by lensing magnification, we do not incorporate lensing corrections into the results of our SED-fitting analysis.

\section{Results and Discussion}\label{sec:results}
\subsection{Results}\label{subsec:results}

\begin{figure*}[htbp]
\centering
\begin{tabular}{c *{3}{c}}
\raisebox{0.2\height}{\rotatebox{90}{\shortstack{All components rising}}} & \multicolumn{2}{c}{\includegraphics[width=0.95\textwidth]{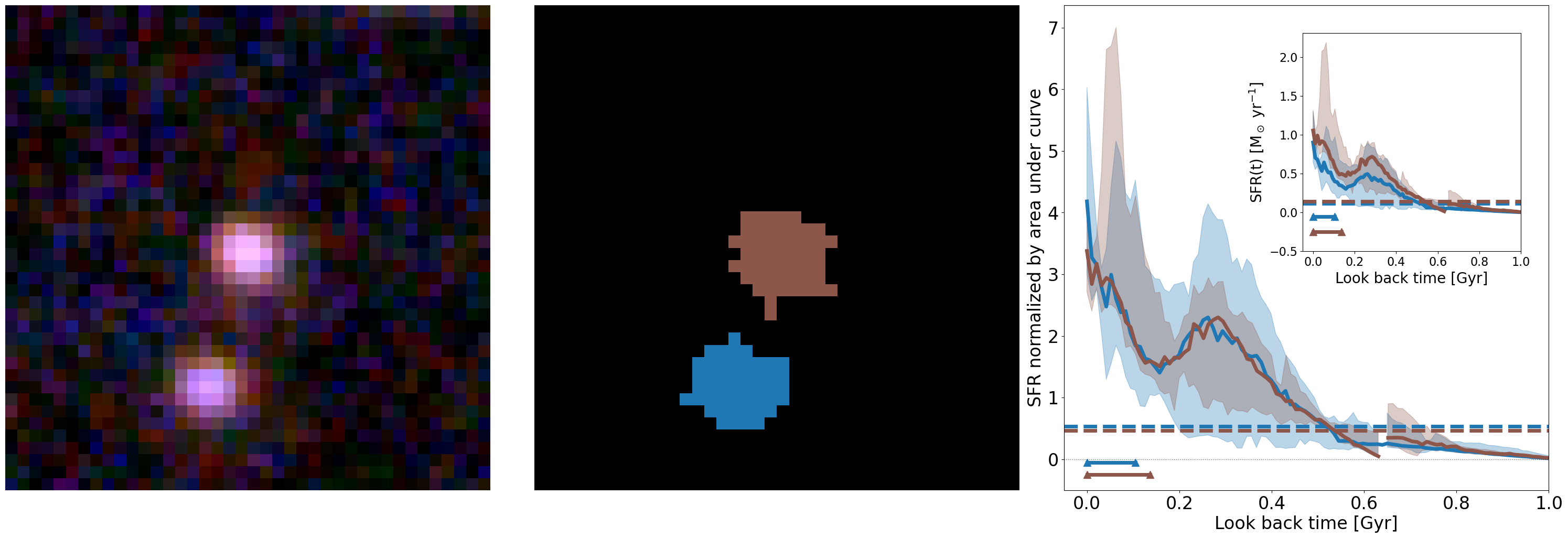}} \\
\raisebox{0.4\height}{\rotatebox{90}{\shortstack{All components declining}}} & \multicolumn{2}{c}{\includegraphics[width=0.95\textwidth]{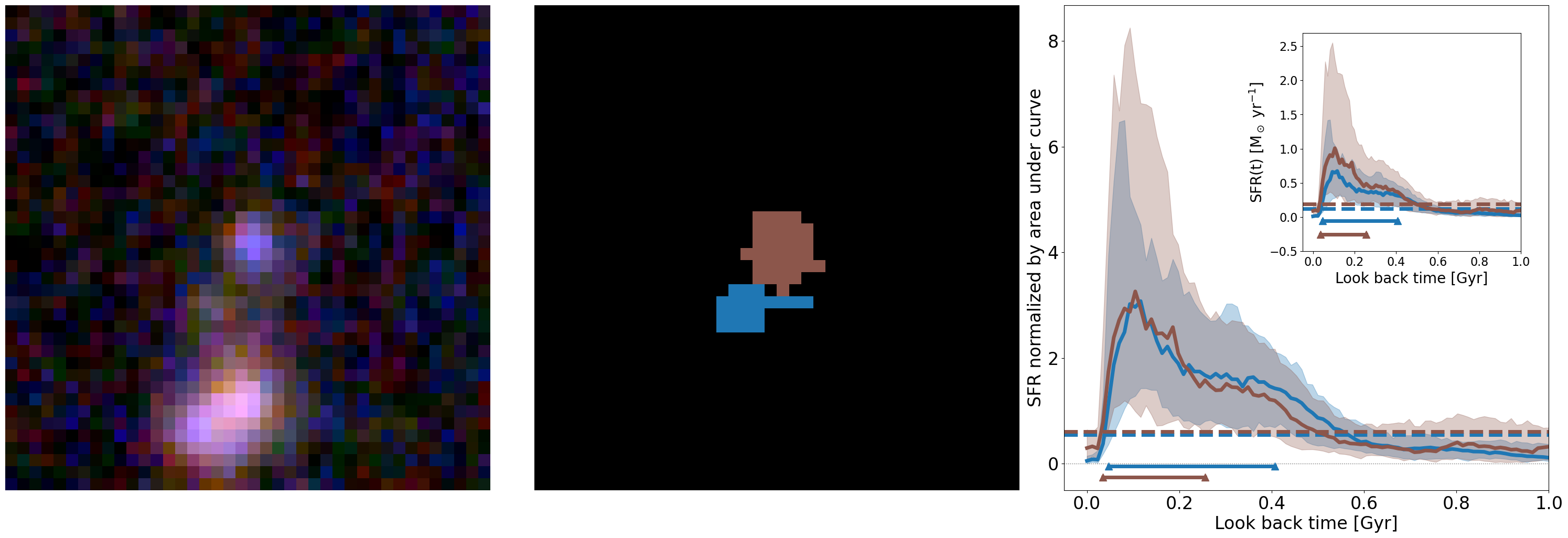}}\\
\raisebox{0.4\height}{\rotatebox{90}{\shortstack{Mixed components}}} & \multicolumn{2}{c}{\includegraphics[width=0.95\textwidth]{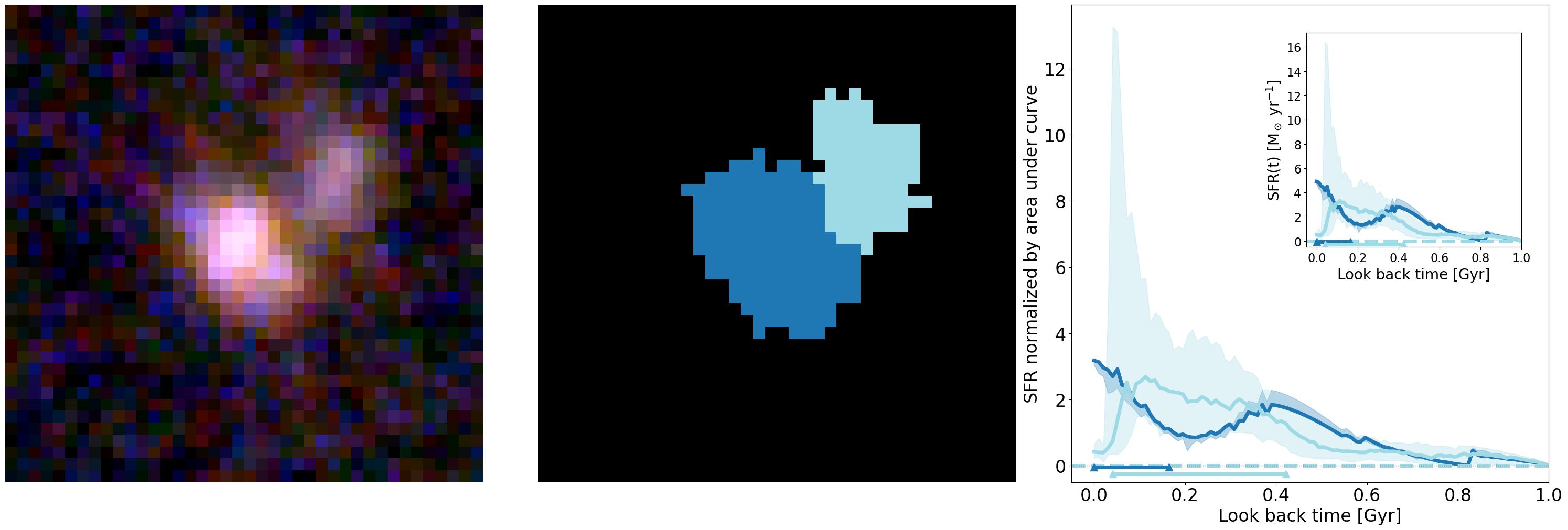}}\\
\end{tabular}
\caption{Examples of star formation histories in our multi-component systems. From top to bottom, {Row 1:} multi-component system with all components rising star formation, {Row 2:} multi-component system with all components declining star formation, {Row 3:} multi-component system with a mixture of rising and declining components. From left to right: RGB image, \textsc{photutils} segmentation map, the quantitative SFH curves of each component normalized to the area under each curve, with the inset figure showing the actual SFHs by $M_{*}$/yr$^{-1}$. Curve colors are matched to the component colors in the segmentation maps. The solid horizontal lines connecting the triangles show the burst duration, $\Delta t_{\rm{burst}}$, for each component, and the dashed horizontal lines represents the {average} SFR outside of the burst
, with line colors matched to the component colors.}
\label{fig:SFHs}
\end{figure*}

 With each system deblended into its components, we examine the star formation histories of each component obtained with \textsc{Dense Basis}  in Sec.~\ref{SEDFIT} and further sub-classify each component based on its SFH.  We find that the 21 systems fall into three categories: systems in which \textbf{a)} all components have rising SFHs at the time of observation, \textbf{b)} systems in which all components have {finished their phase of rising SFR and now have} declining SFHs, and \textbf{c)} systems with a mixture of rising and declining components. Figure \ref{fig:SFHs} shows examples of each type of system, while the type of component mix is indicated for all 21 systems within the panels of Fig.~\ref{fig:combined_mosaics}.
 
 We next find that each of our 58 total components fit into one of the following categories, with the number of components in each group in parentheses: 
\begin{itemize}
\item components {where the }SFR {is} in an increasing state (`rising') (19 components)
\item components {where the SFR has finished its increasing state and is now in a} decreasing state (`declining') (39 components)
\end{itemize}
We also note that none of our components have a flat SFH, where the SFH remains (approximately) constant  over its entire lifetime. All appear to have recent bursts of star formation, albeit (as mentioned above) some of these are already declining while others are still rising. 

The notable absence of components with flat SFHs {is likely not an indicator of a real absence of such systems in the high-$z$ universe. Instead, we attribute their absence from our sample to} \textbf{a)} the difficulty of detecting high redshift systems with little to no recent star formation, \textbf{b)} low SFR systems not being selected for spectroscopic followup (it can be seen in Fig. \ref{fig:ms} that all of our objects lie above the main sequence), \textbf{c)} objects without strong emission lines being less likely to secure spectroscopic redshifts, and \textbf{d)} possible smoothing of SFHs by \textsc{Dense Basis}. A more detailed investigation of this topic is beyond the scope of this work; we note this potential bias and proceed to focus on the objects in our sample.

For each of the component SFHs we define a `star formation burst timeframe' by first identifying the epoch of peak star formation ($t_{\mathrm{peak}}$). Next, we define the burst boundaries $t_{\mathrm{burst,start}}$ (time the burst starts), and $t_{\mathrm{burst,end}}$ (time the burst ends) by extending the SFH outward from $t_{\mathrm{peak}}$ until the following conditions are met:
\begin{enumerate}
    \item The SFR drops below $50\%$ of the SFR at $t_{\mathrm{peak}}$

    \item The gradient, or the absolute rate of change of SFR ($|d\mathrm{SFR}(t)/dt|$) falls below $50\%$ of the maximum slope ($|d\mathrm{SFR}(t)/dt|_{\max}$) recorded during the burst’s rise or decline.

\end{enumerate}
{We choose $50\%$ as our threshold instead of a lower percentage, as this value most consistently captures the full burst timeframe across all systems.} In the case that the SFR or its gradient remains above these thresholds at the limits of the SFH curve (e.g., at zero lookback time or $t=0$), the boundary is truncated at the edge of the observed curve. This approach, which uses both the intensity of the SFR and the gradient, filters out gradual increases in star formation and better identifies the boundaries of star forming bursts. For each component, the time between the start and end of the burst is $\Delta t_\mathrm{burst} = t_{\mathrm{burst,end}}-t_{\mathrm{burst,start}}$ and is indicated by the horizontal bars at the bottom of the SFH panels in Fig.~\ref{fig:SFHs}.  The median burst duration is $0.18\pm0.05$ Gyr, which is short compared to the $\sim$1~Gyr age of the universe at the redshift of objects{, and also consistent with the findings in \citet{2008A&A...492...31D}}. 
Notably, when the SFHs are in the declining mode, these SFH declines are rapid, as can be seen in the middle and bottom rows of in Fig.~\ref{fig:SFHs}.

\subsection{Discussion}\label{sec:discussion}


\subsubsection{Interaction-triggered star formation at $z\sim5$}\label{sec:mergermass}

Of the 48 objects in our parent sample, we identify 21 to be multi-component systems, with similar morphologies as the interacting system {candidates} reported at this redshift in \citet{2024MNRAS.52711372A}. This amounts to $44\%$, which is consistent with the fraction of potential major merger systems found in works at $z\sim5$ such as
\citet{2025MNRAS.544.1412S,2025MNRAS.540..774D, 2025MNRAS.540.2146P}, and in line with the fraction of multi-component systems reported at $z\sim5$ in 
\citet{2009MNRAS.397..208C,2016MNRAS.457..440C,2024MNRAS.52711372A,2025arXiv250814972D}, but less than the $70\%$ reported by \citet{2025ApJ...980..138H}.
We {interpret these systems to be interacting candidates instead of single galaxies with star-forming clumps, as they have visible merger signatures (e.g. tidal tails, bridges, distinct nuclei) in either their RGB images or individual filter images.} We also note that in most of our systems, the second most massive components have masses that lie between 1:1 and 1:4 of the mass of the most massive component (blue points in the middle panel of Fig. \ref{fig:ms}) suggesting that the majority of our {interacting system candidates} are similar-mass interacting systems, {with a caveat that we assume the components are affected equally be lensing.}. Many of our {interacting system candidates} also have additional components (orange points in the middle panel of Fig. \ref{fig:ms}), many of which have much lower relative masses; these components could be tidal dwarf galaxies, perhaps similar to those seen in high-z merger simulations of \citet{2024ApJ...975..238N}. We leave the detailed investigation{, such as kinematic analysis to confirm that they are indeed mergers,} of these low-mass components to future work. 

As mentioned in Sec.~\ref{sec:results} we find several patterns of SFHs in the components of our multi-component system {candidates}: systems where SFR is rising in all the components, systems where SFR is declining in all the components, and systems with both rising and declining components.
The close proximity on the sky of the components within each system ($\lesssim$5 kpc) suggests that each system is either a galaxy with complex morphology, or that the components are individual galaxies that are caught in a galaxy-galaxy interaction. While several high-$z$ objects have been reported to contain embedded star-forming clumps or star clusters \citep[e.g.,][]{2024Natur.636..332M, 2025NatAs.tmp..157F},  the typical stellar masses of the primary and secondary components in our systems ($\gtrsim 10^{7.5} M_\odot$) are higher than those reported by these authors, making our components unlikely to be such sub-galactic clumps.  Note that while the component masses in our lowest-mass system, the highly-magnified ID 1116665 (row 1, column 1 in Fig. \ref{fig:combined_mosaics}), are in the star cluster mass range, this system is known from spectroscopic kinematics to consist of an interacting galaxy pair (Y.~Asada et al., in prep.). 
{Additionally, if we consider the kpc-scale projected separations in our systems in conjunction with their short duration star formation enhancements \citep{2008A&A...492...31D}, it is likely that these systems are experiencing massive gas inflows driven by the loss of angular momentum via tidal torques \citep{1996ApJ...464..641M}. Because these starbursts rely on a prolonged interaction window to allow these torques to operate, it dynamically implies a low relative velocity ($<100$ km/s), where pair systems are found with enhanced star formation \citep{2003MNRAS.346.1189L,2004MNRAS.352.1081A,2006MNRAS.367.1029S}.}
{\citet{2025arXiv251014743P} studied SFR enhancement in close pair systems, and found the most enhancement in pairs with $\sim5-20$ kpc separation. These enhancements were the most pronounced in SFRs averaged over longer timescales ($\sim50-100$ Myr). The systems in our work, multi-component systems with small separations, have a median burst duration of $0.18\pm0.05$ Gyr. It is feasible that the peaks of these bursts will be detected in the form of longer timescale SFR enhancements.}

These arguments suggest that the galaxy-galaxy interactions interpretation is likely. In what follows, we therefore assume that galay-galaxy interactions are responsible for the bursts of star formation we observe in the SFHs, and ask: what is the contribution of these interaction-induced bursts of star formation to the stellar masses of these systems? 

\subsubsection{Stellar mass formed in the bursts}

We first evaluate the stellar mass growth in each component of our multi-component galaxies during the SF burst using the burst timeframe defined in Section \ref{SEDFIT}. We quantify the mass growth $\gamma$ by taking the ratio of the total mass formed by the end of the burst ($M_{*,\textrm{post-burst}}$), and the mass formed up to the burst ($M_{*,\textrm{pre-burst}}$), with
%
\begin{equation}
\gamma = \frac{M_{*,\textrm{post-burst}}}{M_{*,\textrm{pre-burst}}}
= \frac{\int_{t_{0}}^{t_{\mathrm{burst,end}}} \mathrm{SFR}(t)\, dt }{\int_{t_{{0}}}^{t_{\mathrm{burst,start}}} \mathrm{SFR}(t)\, dt},
\end{equation}
where $t_0$ is the time at the start of the recovered SFH curve, {or in other words the maximum lookback time, determined by \textsc{Dense Basis} based on the input spectroscopic redshift of the galaxy.} 

In addition, we evaluate the ratio of the excess $M_*$, $\varepsilon$, caused by the burst, which we define as the fraction of  additional stellar mass produced above the level expected from the component’s baseline SFR ($\mathrm{SFR}_{\mathrm{base}}$). $\mathrm{SFR}_{\mathrm{base}}$ 
estimated as the {average} SFR outside of the burst timeframe, representing the characteristic pre- and post-burst star formation rate of the component. These $\mathrm{SFR}_{\mathrm{base}}$  values are represented by the dashed horizontal lines in Fig. \ref{fig:SFHs}. 
The excess stellar mass ratio, $\varepsilon$, is then
\begin{equation}
\begin{aligned}
\varepsilon =
\frac{\int_{t_{0}}^{t_{\mathrm{burst, end}}} \mathrm{SFR}(t)\, dt}
{\int_{t_{0}}^{t_{\mathrm{burst, start}}} \mathrm{SFR}(t)\, dt+\mathrm{SFR}_{\mathrm{base}} \, \Delta t_{\mathrm{burst}} },
\end{aligned}
\end{equation}
where $\Delta t_{\mathrm{burst}}= t_{\mathrm{burst, end}} - t_{\mathrm{burst, start}}$ is the duration of the burst.

\begin{deluxetable}{lccc}[ht!]
\tablenum{1}
\tablecaption{Summary of component properties, from left to right, {Column 1:} SFH properties, {Column 2:} number of components in each category, {Column 3:} median mass growth factor $\gamma$ of the component during the burst, {Column 4:} median excess $M_*$ ratio. {Uncertainties represent symmetric 95\% confidence interval ($1.96 \times \text{standard error}$) on the median, derived from 10,000 bootstrap resamples of the component populations.} \label{tab:summary}}

\tabletypesize{\footnotesize}
\tablewidth{0pt}

\tablehead{
\colhead{Type of component} & 
\colhead{\shortstack{\# of\\ components}} & 
\colhead{\shortstack{$\gamma$}} & 
\colhead{\shortstack{$\varepsilon$}}
}

\startdata
Rising SF & 19 & $1.50\pm0.89$ & {$1.22\pm0.41$}\\
Declining SF & 39 & $2.66\pm0.85$ & {$1.58\pm0.25$}\\
\enddata
\end{deluxetable}

The median values of $\gamma$ and $\varepsilon$ for the components of the 21 systems in our sample are summarized in columns 3 and 4 of Table \ref{tab:summary}. We find that for the components with rising star formation, the burst contribution to the component stellar mass has a median $\gamma$ of $1.50\pm0.48$, and the median $\varepsilon$ is {$1.22\pm0.41$}. For the components with declining star formation, we find a median $\gamma$ of $1.66\pm0.43$ during the burst, and the $\varepsilon$ is {$1.58\pm0.25$}.
The fact that the $\gamma$ and $\varepsilon$ values are larger for the declining components than in the rising ones makes sense if we consider that the declining components are at a later stage of the star formation burst than those still in their rising phases. Their larger $\gamma$ and $\varepsilon$ values are a reflection of the fact that the declining components had more time to produce more additional stellar mass during their (more advanced) star formation burst episode. With this in mind we take the values of the declining components as being more informative since they represent the state of the component after the burst is over, or at least mostly over. 

Taking the declining-component values, we  find that the interaction-induced burst activity roughly doubles to triples the stellar mass of the component ($\gamma=2.66\pm0.85$). This suggests that interactions trigger large boosts in star formation, in contrast to what was reported at $4.5<z<6.5$ in \citet{2026MNRAS.546ag008D}. This discrepancy is likely due to our sample primarily consisting of small-separation pairs ($<5$ kpc separation), a level of separation not sampled in \citet{2026MNRAS.546ag008D}. By surveying interacting systems at small separations, interaction-induced star formation has been revealed. 
 We further note that the $z\sim5$ boost in stellar mass is also greater than the merger-induced stellar mass growth of $\sim10-20\%$ reported in lower-$z$ studies by \citet{2013MNRAS.429L..40K,2025MNRAS.538L..31F,2025arXiv251121512E}, suggesting that interactions are more efficient at triggering stellar mass growth at $z\sim5$ than at low redshift, also shown in Fig.~\ref{fig:smar}. Such larger efficiency could, for example, be related to larger gas fractions in high-$z$ galaxies than in galaxies at lower redshifts.

\subsubsection{Total mass growth due to mergers}

In this section, we will assume that our interacting systems naturally lead to mergers and evaluate the total mass growth caused by merger activity at $z\sim5$.
While previous work by \citet{2025MNRAS.540..774D} has considered the mass growth of high-$z$ galaxies due to the combination of pre-existing masses of the merging galaxies (the merger's `dry merger' contribution), we will for the first time also include the stellar mass formed during the galaxy-galaxy interactions (a `wet merger' contribution). We will do so by building on the measurements and formalism presented by \citet{2025MNRAS.540..774D} but will extend it to include the additional new stellar mass formed in the interaction.

\citet{2025MNRAS.540..774D} report that galaxy mass can grow by a factor of $2.77\pm0.99$ between $z=10.5$ and $z=5$ from mass accretion due to major mergers. They also report that at $z\sim5$, stellar mass accretion from major mergers is $\sim 27^{+2}_{-1}\%$ of stellar mass created from intrinsic star formation, and that accretion from major mergers accounts for $21^{+12}_{-6}\%$ of a galaxy's total mass growth.  Studies by 
\citet{2025MNRAS.540.2146P,2026arXiv260218068C} report somewhat smaller contributions of mergers to stellar mass growth, at $\sim10\%$. However, none of these studies considered the effect of the bursts of star formation seen in close galaxy pairs (Sec.~\ref{sec:results}) and are thus based on only the accreted pre-existing `dry merger component' mass from the secondary galaxy. A merger event has in fact two components contributing to stellar mass growth: \textbf{a)} mass accreted from the secondary galaxy (the `dry component,' or 'dry mode,' of the merger), and \textbf{b)} enhanced star formation induced by gas infall (the merger's `wet component' or 'wet mode'). As such, given the significant star formation occurring in interacting galaxies (Sec.~\ref{sec:mergermass}), the impact on stellar mass growth due to mergers at this cosmic epoch can be higher than previous studies have considered.  

In this work, we quantify the total stellar mass growth including both the `dry component' evaluated by \citet{2025MNRAS.540..774D} and the `wet component' evaluated in the previous sections, expanding on the merger mass growth formalism presented by \citet{2025MNRAS.540..774D} and the references within. As noted previously, the majority of our systems are major merger systems, so the values from \citet{2025MNRAS.540..774D} can be adopted.

 We first quantify the mass growth due to merger-induced star formation. When two galaxies merge, it is expected that both the primary and secondary galaxy experience enhancements in star formation \citep{2024A&A...691A..82C}. In Section \ref{sec:mergermass} we found that the {median} excess mass growth due to merger-induced star formation is $\varepsilon \sim 1.5$ at $z\sim5$, {however we note some caveats to this number in Section \ref{caveats}}. The mass added to a galaxy due to merger-induced star formation from a singular merger event is $M_{*}(\varepsilon-1)$. Consequently, the {rate of total mass added to a galaxy by merger-induced star formation}, or merger-induced mass{ rate} (MIMR) at redshift $z$ is
 \begin{equation}
     \mathrm{{MIMR}}(z)=\mathcal{R}_{\mathrm{M}}(z) \times [M_{*,1}(z)(\varepsilon-1)+M_{*,2}(z)(\varepsilon-1)],
\end{equation}
 with $M_{*,1}(z)$ and $M_{*,2}(z)$ the average stellar mass of the primary and secondary galaxy at time $z$ based on galaxy stellar mass functions, adopting from \citet{2025MNRAS.540..774D}. $\mathcal{R}_{\mathrm{M}}(z)$ is the merger rate per galaxy at time $z$, as in \citet{2019ApJ...876..110D,2020ApJ...895..115F,2022ApJ...940..168C,2024A&A...691A..82C}. This makes the specific MIMR (sMIMR, MIMR divided by the average stellar mass of a primary galaxy),
\begin{equation}
     \mathrm{s{MIMR}}(z)=\frac{\mathcal{R}_{\mathrm{M}}(z) \times [M_{*,1}(z)(\varepsilon-1)+M_{*,2}(z)(\varepsilon-1)]}{M_{*,1}(z)}.
\end{equation}
Combining this with the specific mass accretion rate, sMAR($z$), derived in \citet{2025MNRAS.540..774D}, gives the total mass added by mergers as sMAR$^+$($z$), 
\begin{equation}\label{eq:sMARplus}
\begin{aligned}
     \mathrm{sMAR^+}(z) &= \mathrm{sMAR}(z)+\mathrm{sMIMR}(z)\\
    & = \frac{\mathcal{R}_\mathrm{M}(z) \times  {M}_{*,2}(z)}{ {M}_{*,1}(z)} \\
     & + \frac{\mathcal{R}_{\mathrm{M}} (z)\times [M_{*,1}(z)(\varepsilon-1)+M_{*,2}(z)(\varepsilon-1)]}{M_{*,1}(z)}\\
    &=\frac{\mathcal{R}_\mathrm{M}(z)}{M_{*,1}(z)}[M_{*,1}(z)(\varepsilon-1)+M_{*,2}(z)\varepsilon]\\
    &=\mathrm{sMAR}(z)\varepsilon+\mathcal{R}_{\mathrm{M}}(z)(\varepsilon-1).
\end{aligned}
\end{equation}
Note that sMAR$^+$ includes both the accretion of pre-existing mass (the `dry component' of the merger) and the merger-induced star formation (the merger's `wet component'), with the `plus' indicating that it is the mass accretion rate that's enhanced (by induced star formation) as compared to the formalism of \citet{2025MNRAS.540..774D}.
Note that setting $\varepsilon$ to 1 (the case where no additional star formation is triggered) in Eq.~\ref{eq:sMARplus} recovers the equation for the specific mass accretion rate, sMAR, of \citet{2025MNRAS.540..774D}. Positive values of interaction-induced mass formation, $\varepsilon$, produce an enhancement over and above that dry mode-only result. We are thus expanding on this formalism and dry component (sMAR) measurements presented in that study to derive the total rate, sMAR$^+$.

\begin{figure*}[ht]

    \centering
    \includegraphics[width=\textwidth]{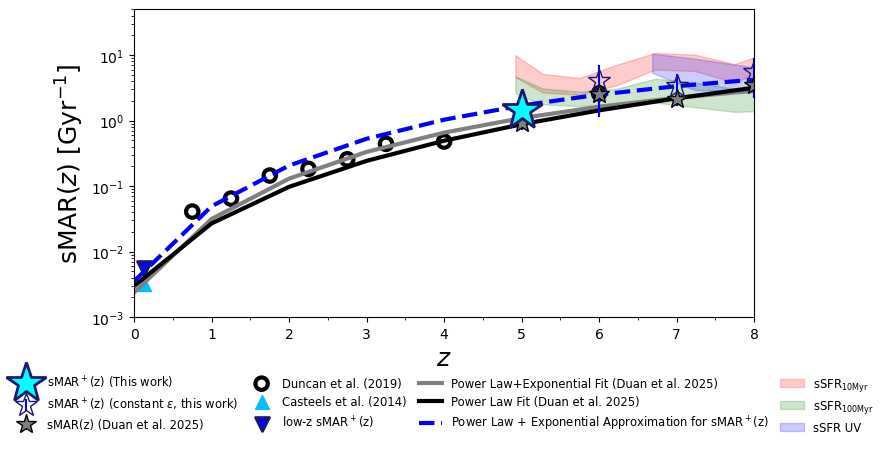}
    \caption{Evolution of the burst-enhanced specific mass accretion rate, sMAR$^+$($z$), with redshift. The total specific mass accretion rate at $z\sim5$, indicated by the blue filled star, is obtained by combining \textbf{a)} sMAR($z$) derived in \citet{2025MNRAS.540..774D}, indicated by the grey stars, with the \textbf{b)} specific merger-induced star formation sMIMR($z$). We also plot sMAR$^+$($z$) at $z\sim6$ and higher, assuming a constant $\varepsilon$ from $z\sim5$.
    Additionally, we derive the value of sMAR$^+$($z$) at low redshifts using our definition, based on sMAR($z$) obtained through \citet{2014MNRAS.445.1157C} indicated by the light blue triangle, and merger induced star formation excess found in \citet{2025arXiv251121512E}. This low-z sMAR$^+$($z$) is indicated by the  blue upside-down triangle.
    We approximate sMAR$^+$($z$) from $0<z<8$ using this low-z point and our high-z points using a power law + exponential fit, shown by the blue line.
    As reference, we include the power law + exponential fit (grey line) and power law fit (black line) to sMAR($z$), as well as specific mass accretion rates at lower redshift based on merger fractions from \citet{2019ApJ...876..110D}, indicated by black rings, all obtained in \citet{2025MNRAS.540..774D}. The shaded areas are intrinsic sSFR from \citet{2025MNRAS.540..774D} on the EPOCHS VI high-redshift sample \citep{2025ApJ...983...30C}, obtained from UV luminosity (blue), and from \textsc{BAGPIPES} \citep{2018MNRAS.480.4379C}  averaging over a 100 Myr timescale (green), and a 10 Myr timescale (red).}
    \label{fig:smar}
\end{figure*}

 We show our value for sMAR$^+$ at $z\sim5$ and compare it with the (dry-only, by construction) \citet{2025MNRAS.540..774D} sMAR($z$) at $z\sim5$ in Fig.~\ref{fig:smar}. sMAR$^+$ at lower redshifts is included as reference. Our sMAR$^+$ was obtained using the values for sMAR($z$) and $\mathcal{R}_{\mathrm{M}}$($z$) from \citet{2025MNRAS.540..774D}, and $\varepsilon$ from Section \ref{sec:mergermass}. We also apply an SFR correction of $(1-R)$ where $R=0.423$, following \citet{2014ARA&A..52..415M} and \citet{2025MNRAS.540..774D}, which accounts for the fraction of mass returned to the interstellar medium through processes such as supernovae. Low redshift sMAR$^+$ is derived from sMAR($z$) obtained through \citet{2014MNRAS.445.1157C} and merger induced star formation excess found in \citet{2025arXiv251121512E}.
 
 We find that sMAR$^+$ at $z$$\sim$$5$ is  {$1.58\pm0.86\times$} sMAR; in other words, including the `wet component' in the merger mass growth calculation results in a mass growth rate that {can be approximately} two times that of the dry-only results.  We thus conclude that when bursts of star formation caused by interactions are included, high-$z$ galaxies {can} grow twice as fast {or greater} as straightforward pair count statistics would tell us.

 Next, we build on the results presented by \citet{2025MNRAS.540..774D} to compare the contribution of mergers to the mass growth of galaxies with that from secular star formation. 
 Interpolating from the percentages derived in \citet{2025MNRAS.540..774D}, the total stellar mass added by mergers can be up to {$62^{+64}_{-39}\%$} of that from intrinsic star formation, shown by the shaded areas in Fig. \ref{fig:smar}. Up to {$38^{+18}_{-20}\%$} of stellar mass added to a galaxy at $z\sim5$ comes from a merger event, {taking into consideration that the intrinsic star formation indicated by the shaded areas can include merger-induced star formation}. 
 Predictions from \citet{2023MNRAS.518.5323H}, and observational results such as \citet{2019ApJ...876..110D,2025MNRAS.540..774D} have already indicated that mergers are the key mechanism of stellar mass growth at high redshift.
 Our inclusion of the mergers' ``wet component" suggest that stellar mass growth due to mergers {can be} significantly boosted by newly-formed stars and, consequently, exceeds the measurements and predictions of these previous works. 


\subsubsection{Growth relative to galaxy population at $z\sim5$}

{To test how our conclusions depend on the assumed baseline SFR, we evaluated our values for $\varepsilon$ where instead of the average SFR, $\mathrm{SFR}_{\mathrm{base}}$ 
{for each component is estimated using the SFMS derived in \citet{2025arXiv250922871M} at $z\sim5$,  $\log\mathrm{SFR}=0.80\log M_*-6.54$. $\mathrm{SFR}_{\mathrm{base}}$ is then the expected SFR of a galaxy with the same stellar mass as each component.} Using this formulation for $\mathrm{SFR}_{\mathrm{base}}$, we find that for the components with rising star formation, the median $\varepsilon$ is {$1.30\pm0.25$}. For the components with declining star formation, we find that the median $\varepsilon$ is {$1.81\pm0.59$}. 
 For the subsequent computations based on sMAR$^+$, we find it to be {$1.98\pm0.82\times$} sMAR. The total stellar mass added by mergers can be up to {$78^{+121}_{-72}\%$} of that from intrinsic star formation, and up to {$44^{+23}_{-38}\%$} of stellar mass added to a galaxy at $z\sim5$ comes from a merger event. As such, the contribution to stellar mass growth by mergers is consistent regardless of our definition of $\mathrm{SFR}_{\mathrm{base}}$.}

\subsection{Caveats}
\label{caveats}
We note a number of caveats with this work. First, as mentioned in Section \ref{sec:Data}, our component definition is based on \textsc{photutils} segmentation maps created from detection maps. This algorithm searches for sources with a method combining multi-thresholding and watershed, and relies on the presence of local peaks and saddles in the flux distribution in the detection maps. As such, if there is a lack of flux peak in a substructure or component, such as low luminosity components \citep{2017A&A...608A..16R}, or if the saddle between flux peaks of potential components is small, the component may not be deblended accurately and therefore missed. Additionally, some galaxies classified as single-component systems may in reality host unresolved or low-luminosity substructure. For these reasons, the single-component galaxies may in reality be multi-component merger galaxies, or recently merged post-coalescence galaxies, and not necessarily be a representative sample of isolated galaxies. 

We also note the existence of possible selection effects in our sample.
As shown in Fig. \ref{fig:ms}, the spectroscopically selected sample mainly consists of galaxies with enhanced SFRs, and is biased towards galaxies and components that are experiencing or have recently ($<$100 Myr) experienced bursts of star formation, which may make their properties not a representative sample of galaxies at this redshift. Galaxies with very faint components with declining or quenched star formation are likely underrepresented in this sample,
which may further affect our statistical results.

For the identification of starbursts from our non-parametric SFHs, they can be subject to uncertainties related to resolution and smoothing from SED-based SFH recovery. Smoothing over short-timescale fluctuations can affect the timing and amplitude of local minima and maxima in the recovered SFH. {In addition, the uncertainties in our calculations in the previous Section are based on scatter in the population, and not the uncertainty in the SFHs themselves.} As such, the timeframe of the bursts in this work are approximate, and the derived burst durations and mass growth factors may not necessarily be exact. {Further, the burstiness and shape of the SFH can be affected by factors such as the SFH priors and models. For example, the Dirichlet prior being locked to 3 in all but the earliest epoch allows for a smooth SFH where the starburst activity can be pushed to a later time.}

 Finally, we only have two strongly constrained points for sMAR$^+$($z$), at $z\sim0$ based on the values from \citet{2014MNRAS.445.1157C} and \citet{2025MNRAS.538L..31F, 2025arXiv251121512E}, and at $z\sim5$ based on our investigation. {In addition, due to our galaxy sample being biased to systems with strong emission lines, it may miss systems where H$\alpha$ is missing, such as dry mergers. As such, our values are likely to be upper limits.} Future works should uncover if $\varepsilon(z)$ changes in earlier epochs, and provide better constraints on sMAR$^+$($z$).


\section{Conclusion}
\label{sec:conclusion}

In this work, we studied the star formation histories of 21 close-separation ($\lesssim$~5~kpc) multi-component system {candidates} at $5.0<z_{spec}<5.6$ that we visually identified from a spectroscopically-confirmed parent sample of 48 galaxies at these redshifts. A key element of our study is that at these redshifts our NIRCam F410M imaging captures the $H\alpha$ emission line, letting us probe star formation on short timescales in all components of each system irrespective of NIRSpec microshutter placement. 

The main findings of our study of this sample are: 
   \begin{enumerate}

       \item After conducting SED fitting on each component,
       we find that all show recent or ongoing bursts of star formation, which we interpret as being triggered by galaxy-galaxy interactions. In some systems all components have increasing star formation histories, in some all have declining SFHs, and in some there is a mixture of the two behaviors. 

       \item Considering the components with declining SFHs as already having mostly completed their burst, we use their properties to estimate the stellar masses produced during the bursts. We find that the star-forming bursts approximately double the stellar mass of each component and produces a stellar mass excess of $\sim1.58\pm0.86\times$ compared to if there was no burst.
       
       \item Assuming that the components of our multi-component systems represent early stages of galaxy-galaxy mergers, we find that mergers account for a combined {$\sim38^{+18}_{-20}\%$} of stellar mass growth in a typical $z\sim5$ galaxy, with $\sim21^{+12}_{-6}\%$ coming from the accretion of pre-existing mass (`dry component' of the merger) and {$\sim17^{+18}_{-23}\%$} coming from merger-induced star formation (the merger's `wet component'). Intrinsic star formation accounts for the remaining $\sim62\%$ of mass growth, indicating that mergers can account for approximately {half of} the stellar mass growth in high-$z$ galaxies.

   \end{enumerate}

This study suggests that interactions and mergers play a very significant role in building galaxies during the first Gyr of cosmic time, with significant amount of new stars apparently being created in interaction-induced bursts of star formation occurring during the merger process. However, further work, such as through kinematics analyses, will be needed to definitively confirm the nature of the components and our assumption that they represent galaxy-galaxy mergers. Detailed studies of individual systems will also be needed to elucidate the way that interactions trigger star-forming bursts at these epochs. Nevertheless, the present work suggests that both `dry' and `wet' modes of the merger process play important roles in building high-$z$ galaxies. They also  highlight the fact that the boosting of star-formation in high-$z$ mergers, while important in assembling galaxies and intense, is of relatively short duration and consequently can be missed in all but closely-separated galaxy pairs. 


\begin{acknowledgments}

This research was enabled by Canadian Space Agency grants 18JWST-GTO1, 24JWGO3A10, 23JWGO2A13, 23JWGO2B15 and 24JWGO2A04 and Natural Sciences and Engineering Research Council (NSERC) of Canada grants RGPIN-2020-06023, RGPAS-2020-00065, and RGPIN-2026-08047. 

MB acknowledges support from the ERC Grant FIRSTLIGHT \# 101053208, Slovenian national research agency ARIS through grants N1-0238 and P1-0188, and ESA PRODEX Experiment Arrangements No. 4000146646 and 4000149972. DM acknowledges generous support from the Leonard and Jane Holmes Bernstein Professorship in Evolutionary Science. Support for program JWST-GO-03362, provided through a grant from the STScI under NASA contract NAS5-03127, is acknowledged. AM acknowledges support from the Yavin Family Fund.

This research used the Canadian Advanced Network For Astronomy Research (CANFAR) platform operated in partnership by the Canadian Astronomy Data Centre and The Digital Research Alliance of Canada with support from the National Research Council of Canada, the Canadian Space Agency, CANARIE, and the Canada Foundation for Innovation.

This work is based on observations made with the NASA/ESA/CSA James Webb Space Telescope. The data were obtained from the Mikulski Archive for Space Telescopes at the Space Telescope Science Institute, which is operated by the Association of Universities for Research in Astronomy, Inc., under NASA contract NAS 5-03127 for JWST. JWST observations are associated with programs JWST-GTO-1208, -4527, and GO-3362 (doi:10.17909/ph4n-6n76, doi:10.17909/cyh7-mm53,
doi:10.17909/18nv-np70).

\software{astropy \citep{2013A&A...558A..33A,2018AJ....156..123A,2022ApJ...935..167A},  
          \textsc{Dense Basis} \citep{2017ApJ...838..127I,2019ApJ...879..116I},
            \textsc{eazy-py}
          \citep{2021zndo...5012704B},
          \textsc{photutils}
          \citep{larry_bradley_2025_14889440}}

\end{acknowledgments}

\bibliography{references}{}
\bibliographystyle{aasjournalv7}



\end{document}